\documentclass[%
reprint,
superscriptaddress,
amsmath,amssymb,
aps,
]{revtex4-2}

\usepackage{graphicx}
\usepackage{dcolumn}
\usepackage{bm}
\usepackage{multirow}
\usepackage{float}
\usepackage{CJK}
\usepackage{array}
\usepackage{subfigure}
\usepackage{soul}
\usepackage{color}
\soulregister{\onlinecite}7
\soulregister{\cite}7
\soulregister{\ref}7
\sethlcolor{yellow}

\begin{document}

\preprint{APS/123-QED}

\title{Exotic Dark Matter Search with CDEX-10 Experiment
 at China Jinping Underground Laboratory}

\author{W. H. Dai}
\author{L. P. Jia}
\author{H. Ma}
 \email{mahao@tsinghua.edu.cn}
\author{Q. Yue}
 \email{yueq@mail.tsinghua.edu.cn}
\author{K. J. Kang}
\author{Y. J. Li}
\affiliation{Key Laboratory of Particle and Radiation Imaging 
(Ministry of Education) and Department of Engineering Physics, 
Tsinghua University, Beijing 100084}


\author{H. P. An}
\affiliation{Department of Physics, Tsinghua University, Beijing 100084}

\author{Greeshma C.}
\altaffiliation[Participating as a member of ]{TEXONO Collaboration}
\affiliation{Institute of Physics, Academia Sinica, Taipei 11529}

\author{J. P. Chang}
\affiliation{NUCTECH Company, Beijing 100084}

\author{Y. H. Chen}
\affiliation{YaLong River Hydropower Development Company, Chengdu 610051}

\author{J. P. Cheng}
\affiliation{Key Laboratory of Particle and Radiation Imaging 
(Ministry of Education) and Department of Engineering Physics, 
Tsinghua University, Beijing 100084}
\affiliation{College of Nuclear Science and Technology, Beijing Normal University, Beijing 100875}

\author{Z. Deng}
\affiliation{Key Laboratory of Particle and Radiation Imaging 
(Ministry of Education) and Department of Engineering Physics, 
Tsinghua University, Beijing 100084}

\author{C. H. Fang}
\affiliation{College of Physics, Sichuan University, Chengdu 610065}

\author{X. P. Geng}
\affiliation{Key Laboratory of Particle and Radiation Imaging 
(Ministry of Education) and Department of Engineering Physics, 
Tsinghua University, Beijing 100084}

\author{H. Gong}
\affiliation{Key Laboratory of Particle and Radiation Imaging 
(Ministry of Education) and Department of Engineering Physics, 
Tsinghua University, Beijing 100084}

\author{Q. J. Guo}
\affiliation{School of Physics, Peking University, Beijing 100871}

\author{X. Y. Guo}
\affiliation{YaLong River Hydropower Development Company, Chengdu 610051}

\author{L. He}
\affiliation{NUCTECH Company, Beijing 100084}

\author{S. M. He}
\affiliation{YaLong River Hydropower Development Company, Chengdu 610051}

\author{J. W. Hu}
\affiliation{Key Laboratory of Particle and Radiation Imaging 
(Ministry of Education) and Department of Engineering Physics, 
Tsinghua University, Beijing 100084}

\author{H. X. Huang}
\affiliation{Department of Nuclear Physics, China Institute of Atomic Energy, Beijing 102413}

\author{T. C. Huang}
\affiliation{Sino-French Institute of Nuclear and Technology, Sun Yat-sen University, Zhuhai 519082}

\author{H. T. Jia}
\author{X. Jiang}
\affiliation{College of Physics, Sichuan University, Chengdu 610065}

\author{S. Karmakar}
\altaffiliation[Participating as a member of ]{TEXONO Collaboration}
\affiliation{Institute of Physics, Academia Sinica, Taipei 11529}

\author{H. B. Li}
\altaffiliation[Participating as a member of ]{TEXONO Collaboration}
\affiliation{Institute of Physics, Academia Sinica, Taipei 11529}


\author{J. M. Li}
\author{J. Li}
\affiliation{Key Laboratory of Particle and Radiation Imaging 
(Ministry of Education) and Department of Engineering Physics, 
Tsinghua University, Beijing 100084}

\author{Q. Y. Li}
\author{R. M. J. Li}
\affiliation{College of Physics, Sichuan University, Chengdu 610065}


\author{X. Q. Li}
\affiliation{School of Physics, Nankai University, Tianjin 300071}

\author{Y. L. Li}
\affiliation{Key Laboratory of Particle and Radiation Imaging 
(Ministry of Education) and Department of Engineering Physics, 
Tsinghua University, Beijing 100084}

\author{Y. F. Liang}
\affiliation{Key Laboratory of Particle and Radiation Imaging 
(Ministry of Education) and Department of Engineering Physics, 
Tsinghua University, Beijing 100084}

\author{B. Liao}
\affiliation{College of Nuclear Science and Technology, Beijing Normal University, Beijing 100875}

\author{F. K. Lin}
\altaffiliation[Participating as a member of ]{TEXONO Collaboration}
\affiliation{Institute of Physics, Academia Sinica, Taipei 11529}

\author{S. T. Lin}
\author{S. K. Liu}
\affiliation{College of Physics, Sichuan University, Chengdu 610065}

\author{Y. D. Liu}
\affiliation{College of Nuclear Science and Technology, Beijing Normal University, Beijing 100875}

\author{Y. Liu}
\affiliation{College of Physics, Sichuan University, Chengdu 610065}

\author{Y. Y. Liu}
\affiliation{College of Nuclear Science and Technology, Beijing Normal University, Beijing 100875}

\author{Z. Z. Liu}
\affiliation{Key Laboratory of Particle and Radiation Imaging 
(Ministry of Education) and Department of Engineering Physics, 
Tsinghua University, Beijing 100084}

\author{Y. C. Mao}
\affiliation{School of Physics, Peking University, Beijing 100871}

\author{Q. Y. Nie}
\affiliation{Key Laboratory of Particle and Radiation Imaging 
(Ministry of Education) and Department of Engineering Physics, 
Tsinghua University, Beijing 100084}

\author{J. H. Ning}
\affiliation{YaLong River Hydropower Development Company, Chengdu 610051}

\author{H. Pan}
\affiliation{NUCTECH Company, Beijing 100084}

\author{N. C. Qi}
\affiliation{YaLong River Hydropower Development Company, Chengdu 610051}

\author{J. Ren}
\author{X. C. Ruan}
\affiliation{Department of Nuclear Physics, China Institute of Atomic Energy, Beijing 102413}



\author{Z. She}
\affiliation{Key Laboratory of Particle and Radiation Imaging 
(Ministry of Education) and Department of Engineering Physics, 
Tsinghua University, Beijing 100084}

\author{M. K. Singh}
\altaffiliation[Participating as a member of ]{TEXONO Collaboration}
\affiliation{Institute of Physics, Academia Sinica, Taipei 11529}
\affiliation{Department of Physics, Banaras Hindu University, Varanasi 221005}

\author{T. X. Sun}
\affiliation{College of Nuclear Science and Technology, Beijing Normal University, Beijing 100875}

\author{C. J. Tang}
\affiliation{College of Physics, Sichuan University, Chengdu 610065}

\author{W. Y. Tang}
\author{Y. Tian}
\affiliation{Key Laboratory of Particle and Radiation Imaging 
(Ministry of Education) and Department of Engineering Physics, 
Tsinghua University, Beijing 100084}

\author{G. F. Wang}
\affiliation{College of Nuclear Science and Technology, Beijing Normal University, Beijing 100875}

\author{L. Wang}
\affiliation{Department of Physics, Beijing Normal University, Beijing 100875}

\author{Q. Wang}
\author{Y. Wang}
\affiliation{Key Laboratory of Particle and Radiation Imaging 
(Ministry of Education) and Department of Engineering Physics, 
Tsinghua University, Beijing 100084}
\affiliation{Department of Physics, Tsinghua University, Beijing 100084}

\author{Y. X. Wang}
\affiliation{School of Physics, Peking University, Beijing 100871}

\author{H. T. Wong}
\altaffiliation[Participating as a member of ]{TEXONO Collaboration}
\affiliation{Institute of Physics, Academia Sinica, Taipei 11529}

\author{S. Y. Wu}
\affiliation{YaLong River Hydropower Development Company, Chengdu 610051}

\author{Y. C. Wu}
\affiliation{Key Laboratory of Particle and Radiation Imaging 
(Ministry of Education) and Department of Engineering Physics, 
Tsinghua University, Beijing 100084}

\author{H. Y. Xing}
\affiliation{College of Physics, Sichuan University, Chengdu 610065}

\author{R. Xu}
\affiliation{Key Laboratory of Particle and Radiation Imaging 
(Ministry of Education) and Department of Engineering Physics, 
Tsinghua University, Beijing 100084}

\author{Y. Xu}
\affiliation{School of Physics, Nankai University, Tianjin 300071}

\author{T. Xue}
\affiliation{Key Laboratory of Particle and Radiation Imaging 
(Ministry of Education) and Department of Engineering Physics, 
Tsinghua University, Beijing 100084}

\author{Y. L. Yan}
\affiliation{College of Physics, Sichuan University, Chengdu 610065}

\author{L. T. Yang}
\affiliation{Key Laboratory of Particle and Radiation Imaging 
(Ministry of Education) and Department of Engineering Physics, 
Tsinghua University, Beijing 100084}


\author{N. Yi}
\affiliation{Key Laboratory of Particle and Radiation Imaging 
(Ministry of Education) and Department of Engineering Physics, 
Tsinghua University, Beijing 100084}

\author{C. X. Yu}
\affiliation{School of Physics, Nankai University, Tianjin 300071}

\author{H. J. Yu}
\affiliation{NUCTECH Company, Beijing 100084}

\author{J. F. Yue}
\affiliation{YaLong River Hydropower Development Company, Chengdu 610051}

\author{M. Zeng}
\author{Z. Zeng}
\author{B. T. Zhang}
\affiliation{Key Laboratory of Particle and Radiation Imaging 
(Ministry of Education) and Department of Engineering Physics, 
Tsinghua University, Beijing 100084}

\author{F. S. Zhang}
\affiliation{College of Nuclear Science and Technology, Beijing Normal University, Beijing 100875}

\author{L. Zhang}
\affiliation{College of Physics, Sichuan University, Chengdu 610065}

\author{Z. H. Zhang}
\author{Z. Y. Zhang}
\affiliation{Key Laboratory of Particle and Radiation Imaging 
(Ministry of Education) and Department of Engineering Physics, 
Tsinghua University, Beijing 100084}

\author{K. K. Zhao}
\affiliation{College of Physics, Sichuan University, Chengdu 610065}

\author{M. G. Zhao}
\affiliation{School of Physics, Nankai University, Tianjin 300071}

\author{J. F. Zhou}
\affiliation{YaLong River Hydropower Development Company, Chengdu 610051}

\author{Z. Y. Zhou}
\affiliation{Department of Nuclear Physics, China Institute of Atomic Energy, Beijing 102413}

\author{J. J. Zhu}
\affiliation{College of Physics, Sichuan University, Chengdu 610065}

\collaboration{CDEX Collaboration}

\date{\today}

\begin{abstract}
A search for exotic dark matter (DM)
in the sub-GeV mass range has been 
conducted using 205 kg$\cdot$day 
data taken from a p-type point contact germanium detector 
of CDEX-10 experiment 
at China Jinping underground laboratory. 
New low-mass dark matter searching channels,
neutral current fermionic DM absorption 
($\chi+A\rightarrow \nu+A$)
and DM-nucleus 3$\rightarrow$2 scattering 
($\chi+\chi+A\rightarrow \phi+A$),
have been analyzed with an energy threshold of 160 eVee. 
No significant signal was found, 
thus new limits on the DM-nucleon interaction cross section 
are set for both models 
at sub-GeV DM mass region. 
A cross section limit for the fermionic DM absorption 
is set to be $\rm 2.5\times 10^{-46} cm^2$(90\% C.L.)
at DM mass of 10 MeV/c$^2$.
For the DM-nucleus 3$\rightarrow$2 
scattering scenario,
limits are extended to DM mass of 
5 MeV/c$^2$ and 14 MeV/c$^2$
for the massless dark photon and 
bound DM final state, respectively.
\begin{description}
\item[Key words]
sub-GeV dark matter, fermionic dark matter absorption, 
DM-nucleus 3$\rightarrow$2 scattering, CJPL. 
\end{description}
\end{abstract}

\maketitle


\textit{Introduction-}
Various evidence from cosmological and astronomical 
observations strongly supports the existence of 
dark matter (DM) in our Universe 
[\onlinecite{bib:1}]. 
The weakly interacting massive particle (WIMP) 
DM candidate has been searched 
for several decades 
[\onlinecite{bib:2}-\onlinecite{bib:3}], 
and the limit on the WIMP-nucleus cross section 
has been pushed near the neutrino floor for 
DM masses around the GeV scale in 
direct detection experiments
[\onlinecite{bib:4}-\onlinecite{bib:9}].

The null results in WIMP search have motivated
building of alternative dark matter models
[\onlinecite{bib:33}-\onlinecite{bib:35}]. 
Recently, two sub-GeV light DM interaction scenarios, 
the neutral current fermionic DM absorption 
and DM-nucleus 3$\rightarrow$2 scattering, 
have been proposed and searched for experimentally
[\onlinecite{bib:10}-\onlinecite{bib:14}].

The fermionic DM ($\chi$) 
may convert to a neutrino ($\nu$) 
upon interaction with a target ($A$) 
via a neutral current absorption process 
($\chi+A\rightarrow \nu+A$) 
[\onlinecite{bib:10},\onlinecite{bib:11}]. 
The absorption generates a monoenergetic signal 
with an energy proportional to the DM mass in 
direct detection experiments. 
Experiments with a low energy threshold, 
low background, and good energy resolution 
can provide strong constraints on the 
interaction cross section. 
In direct detection experiments, 
only two searches have been performed to date. 
The most stringent limit at
15$\sim$125 MeV/c$^2$ DM mass range is given
by PandaX-4T experiment
and the lowest limit on the cross section of  
1.7$\times10^{-50}$ cm$^2$ (90\%C.L.)
is placed at DM mass of 35 MeV/c$^2$
[\onlinecite{bib:12}]. 
The M\textsc{ajorana} D\textsc{emonstrator}
set limits on the fermionic DM absorption
cross section
at the DM mass of 25$\sim$190 MeV/c$^2$
via the high purity germanium detector technology
[\onlinecite{bib:13}].

In recent work, 
another direct detection channel for sub-GeV DM 
via a DM-nucleus 3$\rightarrow$2 
scattering process
($\chi+\chi+A\rightarrow \phi+A$) 
is proposed [\onlinecite{bib:14}]. 
Two DM candidates ($\chi$) may combine into a 
DM composite state or dark radiation ($\phi$) 
upon the interaction with a target ($A$). 
Similar to the fermionic DM absorption, 
a monoenergetic signal is generated in the 
three-body scattering, and 
the energy passed to the nucleon 
is related to the mass of the initial 
and final states of DM. 
The M\textsc{ajorana} D\textsc{emonstrator} 
performed searches for 
DM-nucleus 3$\rightarrow$2 scattering signals
with 52.6 kg$\cdot$year exposure data and 
an analysis threshold of 1 keVee [\onlinecite{bib:13}].

In this letter, 
we report the search results of 
the neutral current fermionic DM absorption 
and DM-nucleus 3$\rightarrow$2 scattering 
using 205 kg$\cdot$day exposure data from the
CDEX-10 experiment with an analysis threshold of 160 eVee.

\textit{CDEX-10 Experiment-}
The CDEX-10 experiment operates a 
10-kg p-type point contact germanium (PPCGe) detector array 
to search for light DM in 
China Jinping Underground Laboratory (CJPL) 
[\onlinecite{bib:8},\onlinecite{bib:15}]. 
The detector array consists of three triple-element 
PPCGe detector strings, C10-A, B, C, respectively. 
The detector array surrounded by a 20 cm high-purity 
oxygen-free copper shielding is directly immersed 
in a liquid nitrogen (LN) cryostat for cooling. 
The LN cryostat and data acquisition (DAQ) systems 
are placed inside a polyethylene room at CJPL-I 
[\onlinecite{bib:16}]. 
The detailed configuration of the experimental setup 
can be found in [\onlinecite{bib:8},\onlinecite{bib:16}].

One of the C10 detectors, C10-B1, 
achieved the lowest analysis threshold (160 eVee) 
during its data taking from 
February 2017 to August 2018 
[\onlinecite{bib:8},\onlinecite{bib:17}]. 
The dead layer thickness of C10-B1 
is evaluated to be (0.88$\pm$0.12) mm 
[\onlinecite{bib:8},\onlinecite{bib:9}].
The dead layer results in a fiducial mass of 939 g and, 
accordingly, an exposure of 205 kg$\cdot$day.

The data processing procedures, 
including energy calibration, pedestal cut, 
physics event selection, 
and bulk or surface event discrimination, 
were discussed in our previous works
[\onlinecite{bib:8},\onlinecite{bib:18}]. 
The dead time of the DAQ system was measured to be 5.7\% 
by random trigger signals. 
Efficiencies of event trigger and event selection
were derived from a $^{137}$Cs source
following the method described in our
previous works [\onlinecite{bib:8},\onlinecite{bib:9}].
The analysis threshold is selected to be 160 eVee,
corresponding to a combined efficiency of (5.7$\pm$1.48)\%.
The 0.16$\sim$4 keVee spectrum 
after all event selections and efficiency correction 
is used for DM analysis.

\textit{Data Analysis-}
A minimum $\chi^2$ method, 
similar to our previous WIMP analysis
[\onlinecite{bib:8},\onlinecite{bib:17},\onlinecite{bib:19}], 
is used to search for the fermionic DM absorption and 
DM-nucleus 3$\rightarrow$2 scattering. 
The $\chi^2$ is constructed as

\begin{equation}
    \chi^2=
    \sum_{i}^{}\frac{\left [n_i-S_i(m_{\chi},\sigma_{\chi})-B_i  \right ]^2}
    {\sigma_{stat,i}^2+\sigma_{syst,i}^2},
    \label{eq:1}
\end{equation}

\noindent where $n_i$ is the measured event rate in $i^{th}$ energy bin, 
$\sigma_{stat,i}$ and $\sigma_{syst,i}$ 
are the statistical and systematic uncertainty, respectively.
The systematic uncertainty includes uncertainties in
exposure, event select efficiency and bulk/surface event discrimination [\onlinecite{bib:8}].
$S_i(m_{\chi},\sigma_{\sigma})$ is the expected 
DM event rate in $i^{th}$ energy bin at 
DM mass m$_{\chi}$ and DM-nucleus interaction cross section $\sigma_{\chi}$. 
$B_i$ is the background event rate in the $i^{th}$ energy bin.

\begin{figure}[htb]
    \centering
    \subfigure
    {
    \includegraphics[width=1.0\hsize]
    {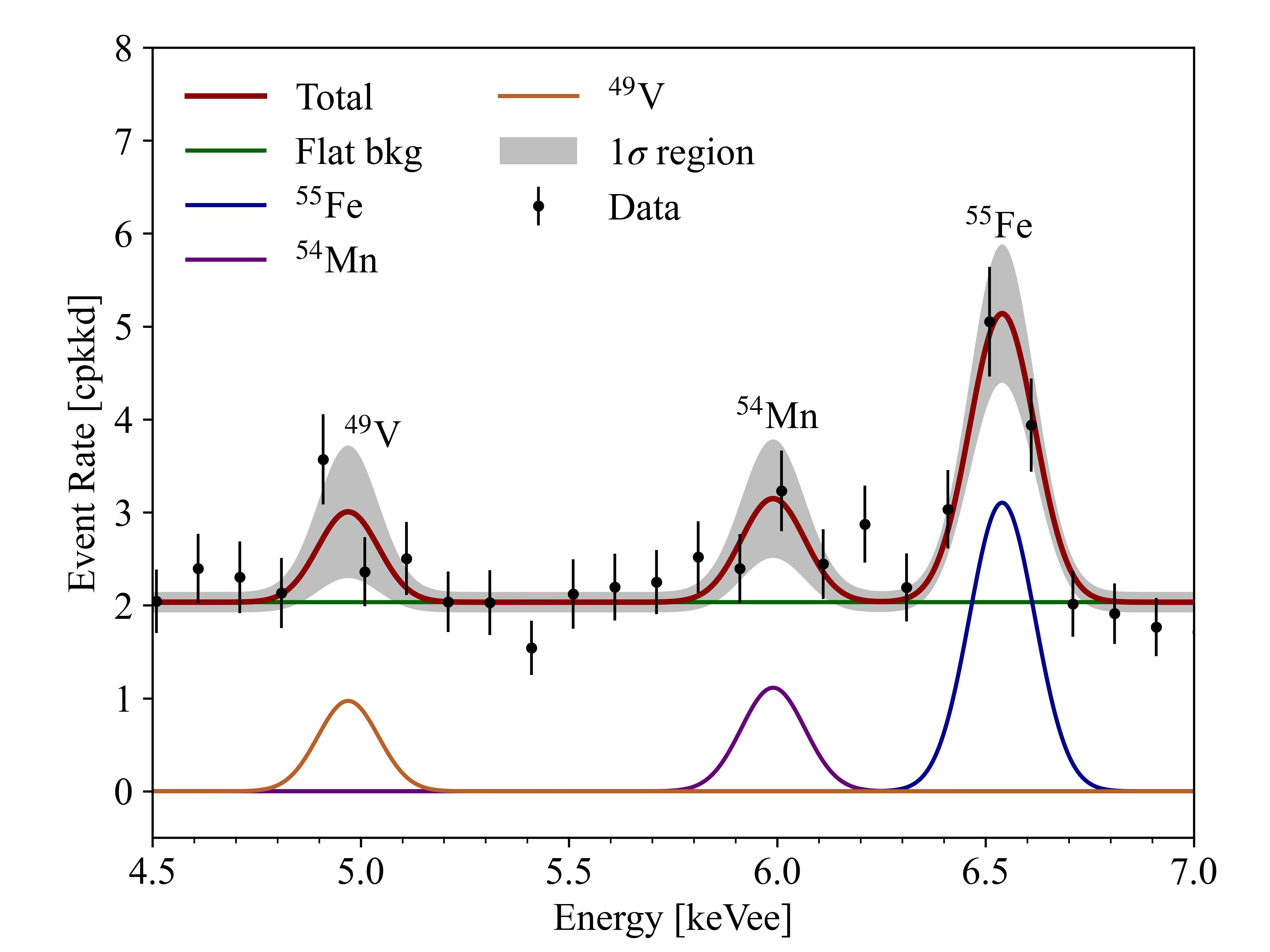}
    }
    \subfigure
    {
    \includegraphics[width=1.0\hsize]
    {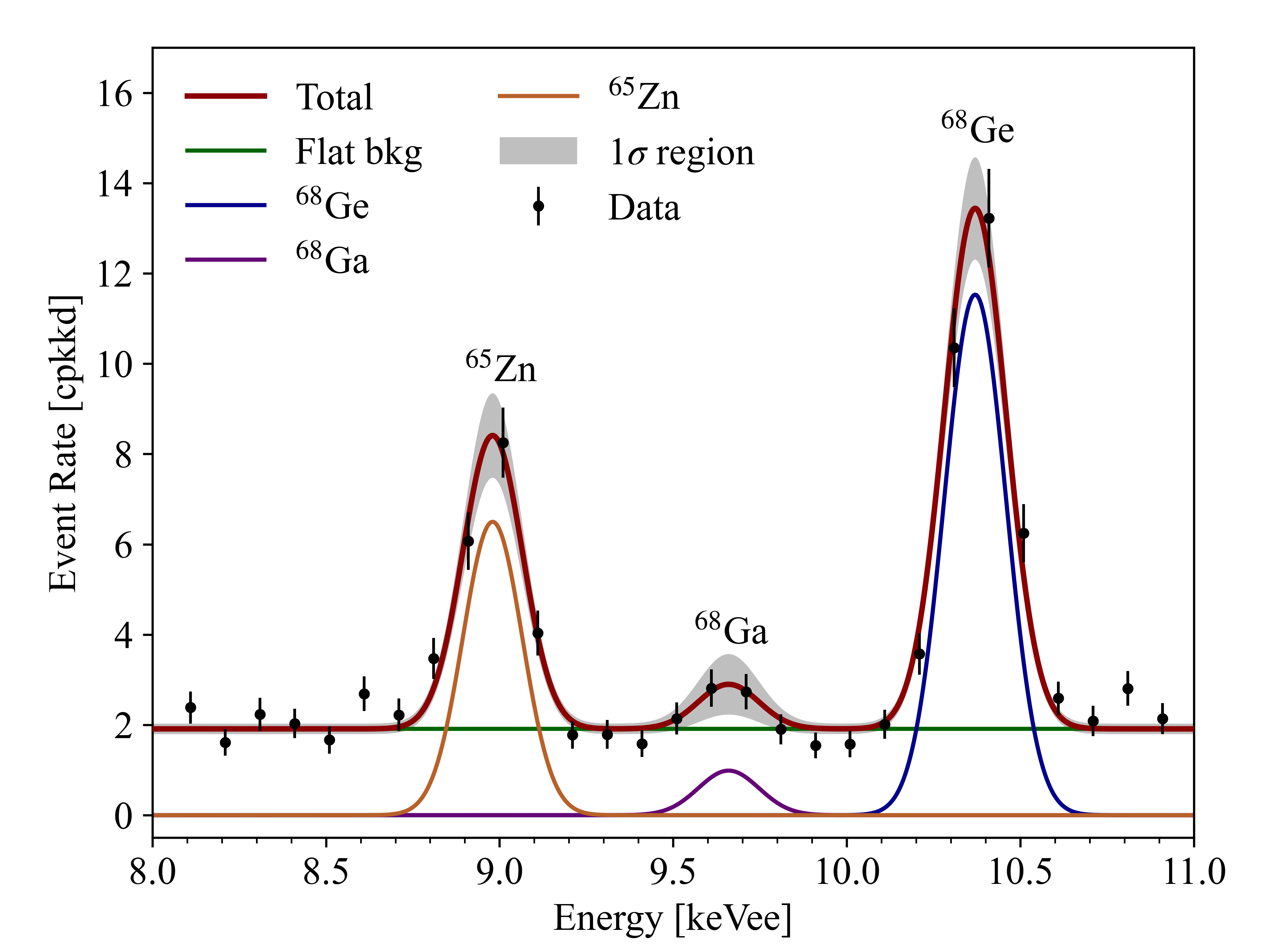}
    }
    \caption{\label{fig:KXFit} Spectra fit for the 
    intensities of the six K-shell X-ray peaks, 
    the 1$\sigma$ uncertainty band derived 
    from the fit uncertainty is labeled in gray.
    The intensity of L/M-shell X-ray peak is constrained
    by the K-shell X-ray peak with the known K/L and K/M ratio 
    [\onlinecite{bib:8},\onlinecite{bib:20}].}
\end{figure}

The background event rate ($B$, in counts per keVee per kg per day) of C10-B1 detector
consists of a flat continuum ($p_0$)
plus L-shell and M-shell X-ray peaks from 
$^{68}$Ge ($E_L=$1.298 keV, $E_M=$0.16 keV), 
$^{68}$Ga ($E_L=$1.194 keV, $E_M=$0.14 keV), 
$^{65}$Zn ($E_L=$1.096 keV), $^{55}$Fe ($E_L=$0.764 keV), 
$^{54}$Mn ($E_L=$0.695 keV), and $^{49}$V ($E_L=$0.564 keV):

\begin{eqnarray}
    B=p_0
    + \sum_{L} I_L\cdot \frac{1}{\sqrt{2\pi}\sigma_L}
    \exp \left (-\frac{(E-E_L)^2}{2\sigma_L^2}  \right ) \\ \nonumber
    + \sum_{M} I_M\cdot \frac{1}{\sqrt{2\pi}\sigma_M}
    \exp \left (-\frac{(E-E_M)^2}{2\sigma_M^2}  \right ),
    \label{eq:2}
\end{eqnarray}

\noindent where $E_L$ and $E_M$ are the energies of the 
L-shell and M-shell X-rays, and $I_L$ and $I_M$ are the
corresponding peak intensities.
$\sigma_L$ and $\sigma_M$ are the 
energy resolutions at the peak energy.
As shown in Fig.\ref{fig:KXFit},
the K-shell X-ray peaks
in the 4$\sim$12 keVee spectrum
are measured to constrain
the intensities of the L/M-shell peaks with
known K/L and K/M ratios
[\onlinecite{bib:8},\onlinecite{bib:20}].
The energy resolution function $\sigma(E)=a+\sqrt{b\cdot E}$ [\onlinecite{bib:27}]
is fitted by resolutions of K-shell X-ray peaks ($a=35.5$ eV, $b=2.8$ eV).
Energy resolutions of L/M-shell X-ray peaks are
constrained to the $\pm3\sigma$ uncertainty region of the 
energy resolution function.

The DM-nucleus interaction cross section is probed by minimizing
the $\chi^2$ value at a certain DM mass, 
and the upper limit at the 90\% confidence level (C.L.) 
is computed using the Feldman-Cousins method 
[\onlinecite{bib:21}].
The nuclear recoil quenching factor of Ge calculated
by the \textsc{trim} software package [\onlinecite{bib:8},\onlinecite{bib:24}$-$\onlinecite{bib:26}]
with a 10\% systematic error is adopted in this analysis
like our previous works [\onlinecite{bib:19},\onlinecite{bib:27}].

\textit{Fermionic DM Absorption-}
As proposed in [\onlinecite{bib:10},\onlinecite{bib:11}], 
the fermionic DM may interact with Ge via a 
neutral current (NC) absorption, 
$\chi+Ge\rightarrow \nu+Ge$,
modeled as a Yukawa-like interaction 
with a bosonic mediator. 
Considering a non-relativistic fermionic DM,
its mass dominates its energy and results in
a momentum transfer $q \simeq m_\chi$ and 
nuclear recoil energy $E_R \simeq m_\chi^2/2M$, 
where $M$ is the mass of the target. 
The differential event rate ($dR_{NC}/dE_R$) 
and total event rate ($R_{NC}$) of the 
neutral current absorption signal 
can be expressed as [\onlinecite{bib:11}]:

\begin{eqnarray}
    \label{eq:7}
    \frac{dR_{NC}}{dE_R}=
    \frac{\rho_\chi }{m_\chi}\cdot \sigma_{NC}\\ \nonumber
    \times\frac{1}{M_T}\sum_{j}^{}N_jM_jA_j^2F_j^2&(m_\chi)
    \frac{\sqrt{2E_RM_j}}{2p_\nu m_\chi^2}
    \left \langle \frac{1}{v_\chi} \right \rangle_{v_\chi> v_{min}},
\end{eqnarray}

\begin{eqnarray}
    R_{NC}=
    \frac{\rho_\chi}{m_\chi}\cdot \sigma_{NC}\cdot
    \frac{1}{M_T}\sum_{j}^{}N_jM_jA_j^2F_j^2(m_\chi),
    \label{eq:8}
\end{eqnarray}

\noindent where $\sigma_{NC}$ is the DM-nucleon 
interaction cross section, 
$m_\chi$ is the DM mass. 
The local DM density $\rho_\chi$ is set to 
0.3 GeV/cm$^3$ as recommended in Ref [\onlinecite{bib:22}]. 
$M_T=\sum_{j}^{}N_jM_j$ 
is the total target mass with 
$N_j$ and $M_j$ corresponding to the number and mass of 
isotope $j$, respectively. 
$F_j(m_\chi)$ is the normalized Helm form factor for 
isotope $j$ evaluated at momentum transfer 
$q=m_\chi$ [\onlinecite{bib:23}]. 
$p_\nu=\sqrt{q(2m_\chi-q-2E_R)}$
denotes the momentum of the outgoing neutrino. 
$v_{min}$ is the minimum DM velocity at a 
given recoil energy $E_R$ [\onlinecite{bib:11}].

\begin{figure}[htb]
    \includegraphics
    [width=1.0\hsize]
    {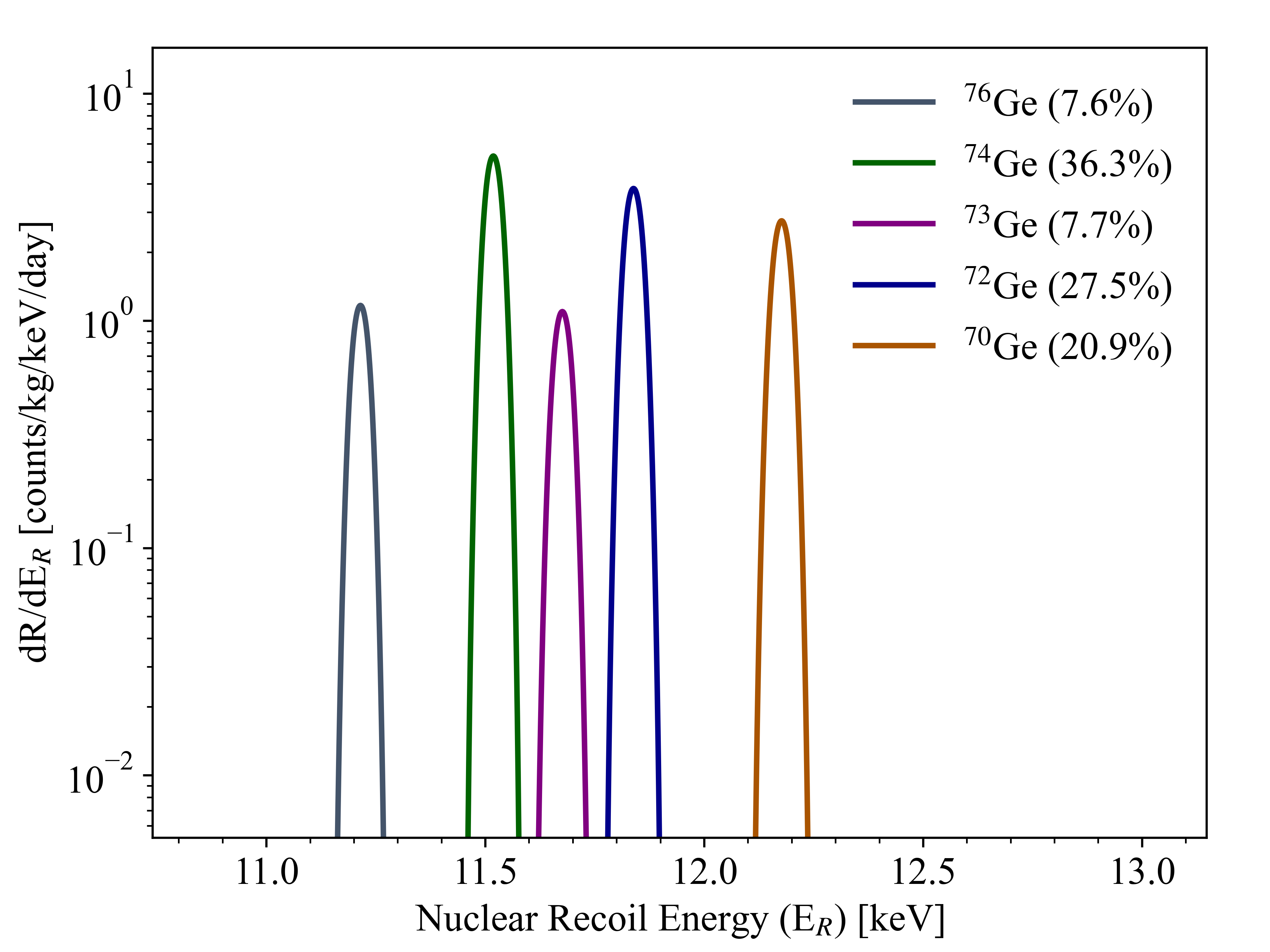}
    \caption{\label{fig:spExpEr} Nuclear recoil energy spectra 
    of the fermionic DM absorption signals for 
    Ge target, with a DM mass 
    $m_\chi$ = 40 MeV/c$^2$ and the
    DM-nucleon cross section $\sigma_{NC}=10^{-45}$ cm$^2$ 
    for all Ge isotopes.}
\end{figure}

\begin{figure}[htb]
    \includegraphics
    [width=1.0\hsize]
    {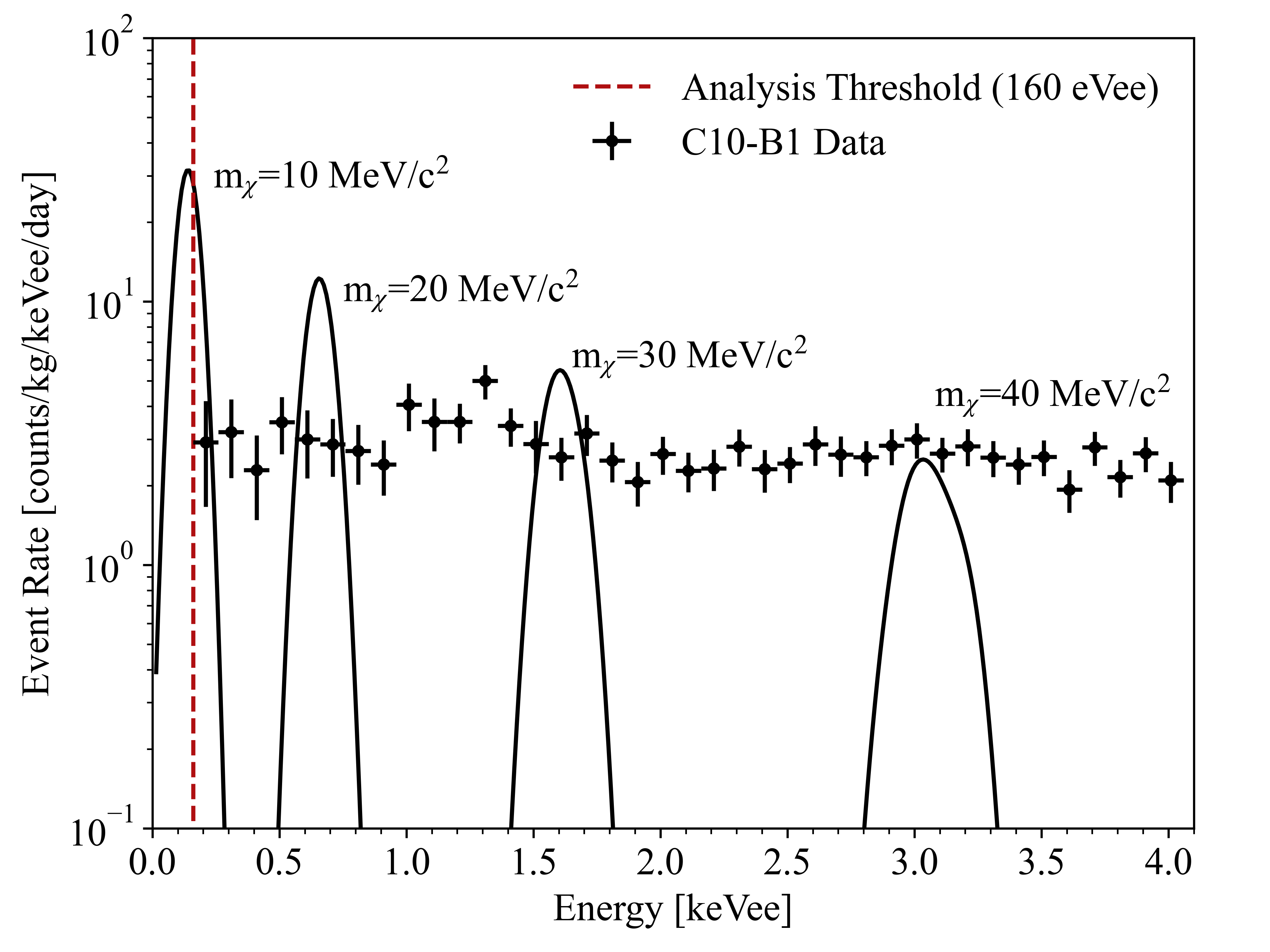}
    \caption{\label{fig:spExp} Expected spectra of the 
    fermionic DM absorption signals for 
    $m_\chi$ = 10, 20, 30, 40 MeV/c$^2$ with a
    DM-nucleon cross section $\sigma_{NC}=10^{-45}$ cm$^2$. 
    The C10-B1 spectrum is shown as the black dots
    with a bin size of 100 eVee.
    The red dash line represents the analysis threshold.}
\end{figure}

As shown in Eq(\ref{eq:7}), the DM signal rate is contributed
by all Ge isotopes,
and their abundance in the natural Ge are 
adopted for the C10-B1 detector: 
$^{76}$Ge (7.6\%), $^{74}$Ge (36.3\%), 
$^{73}$Ge (7.7\%), $^{72}$Ge (27.5\%), 
and $^{70}$Ge (20.9\%). 
Contributions from different Ge isotopes in the 
nuclear recoil energy spectrum are shown in
Fig.\ref{fig:spExpEr}.
Converting the nuclear recoil energy in Eq(\ref{eq:7})
into visible energy and folding the energy resolution 
of the detector, the expected spectra of 
the fermionic DM absorption signals are 
derived for several $m_\chi$ and a certain $\sigma_{NC}$, 
as shown in Fig.\ref{fig:spExp}.

\begin{figure}[htb]
    \includegraphics
    [width=1.0\hsize]
    {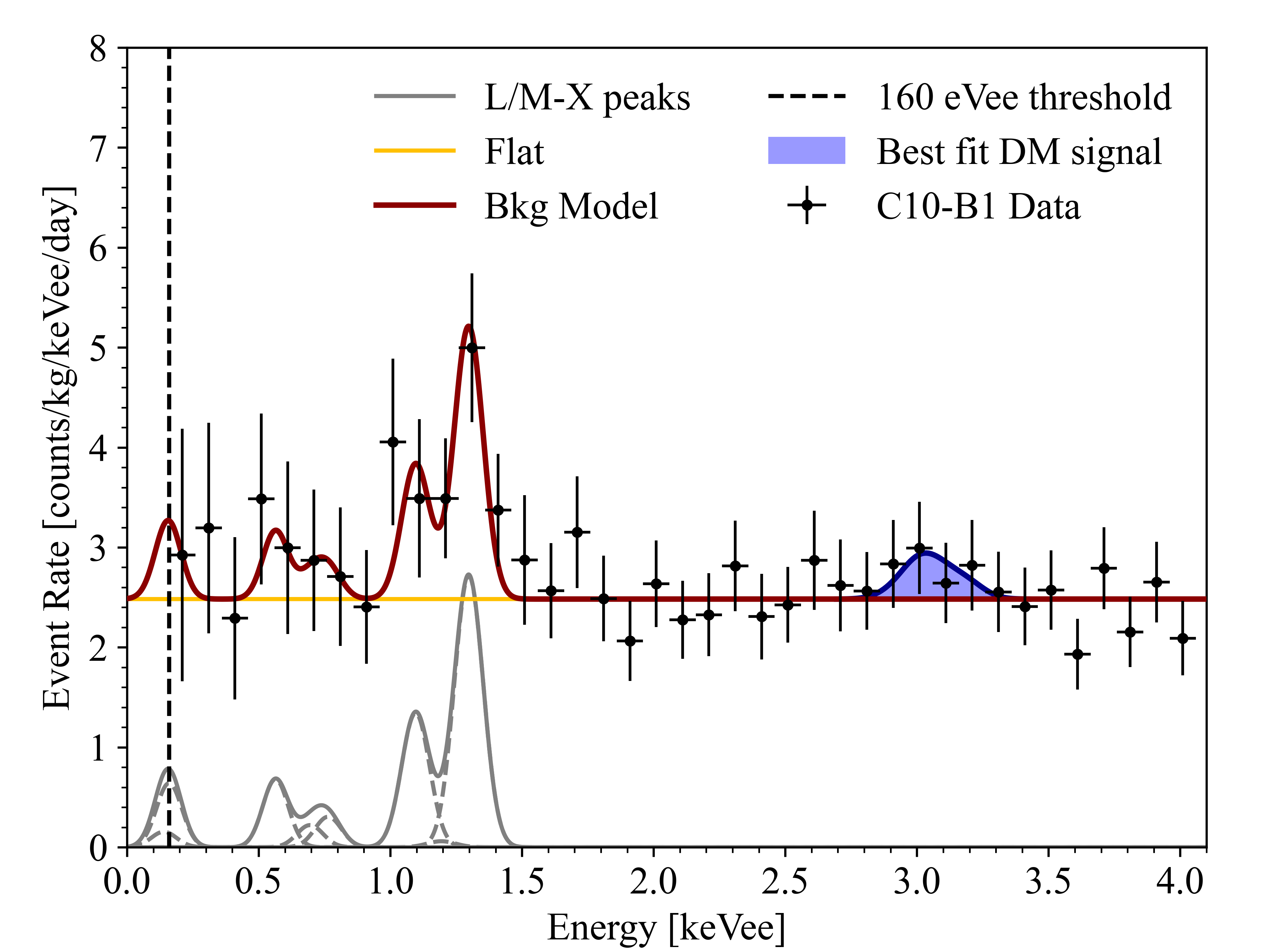}
    \caption{\label{fig:spFit}An example of 
    fermionic DM absorption analysis 
    for $m_\chi$=40 MeV/c$^2$
    with its corresponding best-fit peak signature at 3 keVee.
    The best-fit background model is shown in the red line, 
    the gray lines represent the contributions 
    from L-shell and M-shell X-ray peaks, 
    and the flat background is depicted by the yellow line.
    The best-fit DM signal is labeled in blue.}
\end{figure}

\begin{figure}[htb]
    \includegraphics
    [width=1.0\hsize]
    {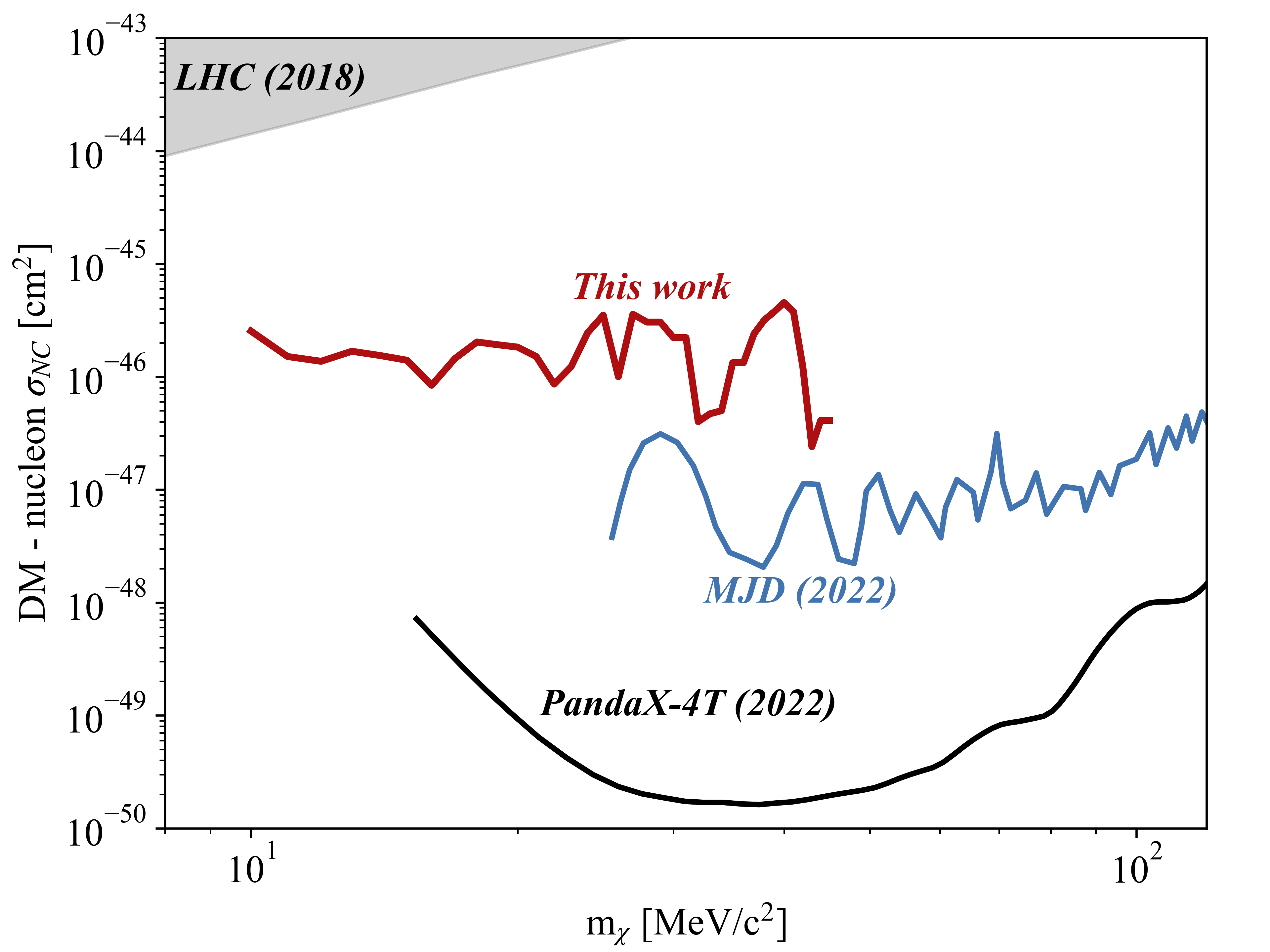}
    \caption{\label{fig:limit-FDM}Upper limit (90\% C.L.) 
    of the fermionic DM absorption cross section $\sigma_{NC}$. 
    The gray shadow region represents the constraints 
    from Z$_0$ monojet searches at the LHC 
    (indirect detection). 
    The result from this work is depicted in solid red,
    results from two other direct detection experiments,
    the M\textsc{ajorana} D\textsc{emonstrator} [\onlinecite{bib:13}]
    (MJD, blue line) 
    and PandaX-4T [\onlinecite{bib:12}] (black line) are superimposed.
    Constraint on $\sigma_{NC}$ for DM mass of $>$10 MeV/c$^2$ is 
    achieved in this work.}
\end{figure}

We scan the DM mass in the range of 10$\sim$45 MeV/c$^2$
(corresponding to 0.16$\sim$4 keVee in the measured energy spectrum), 
and no significant signal is observed. 
The best fit spectrum for $m_\chi\sim$40 MeV/c$^2$ 
and the corresponding best fit DM peak signature at 3 keVee
are displayed in Fig.\ref{fig:spFit}. 
Upper limits of the DM-nucleon cross section 
superimposed with previous results from 
direct detection experiments 
[\onlinecite{bib:12},\onlinecite{bib:13}] 
and Z$_0$ monojet searches at the LHC 
[\onlinecite{bib:28}] 
are illustrated in Fig.\ref{fig:limit-FDM}. 
Note that we take a local DM density ($\rho_{\chi}$) of 0.3 GeV/cm$^3$
recommended in [\onlinecite{bib:22}] other than 0.4 GeV/cm$^3$
used by the M\textsc{ajorana} D\textsc{emonstrator} [\onlinecite{bib:13}],
resulting in more conservative constraints.
The most stringent limit on 
DM-nucleon cross section is 
$\sigma_{NC}<2.4\times10^{-47}$ cm$^2$
at DM mass of 43.0 MeV/c$^2$ in this work. 
The upper limit on cross section is set to 
$\sigma_{NC}<2.5\times10^{-46}$ cm$^2$
at 10 MeV/c$^2$ DM mass.

\begin{figure}[H]
    \includegraphics
    [width=1.0\hsize]
    {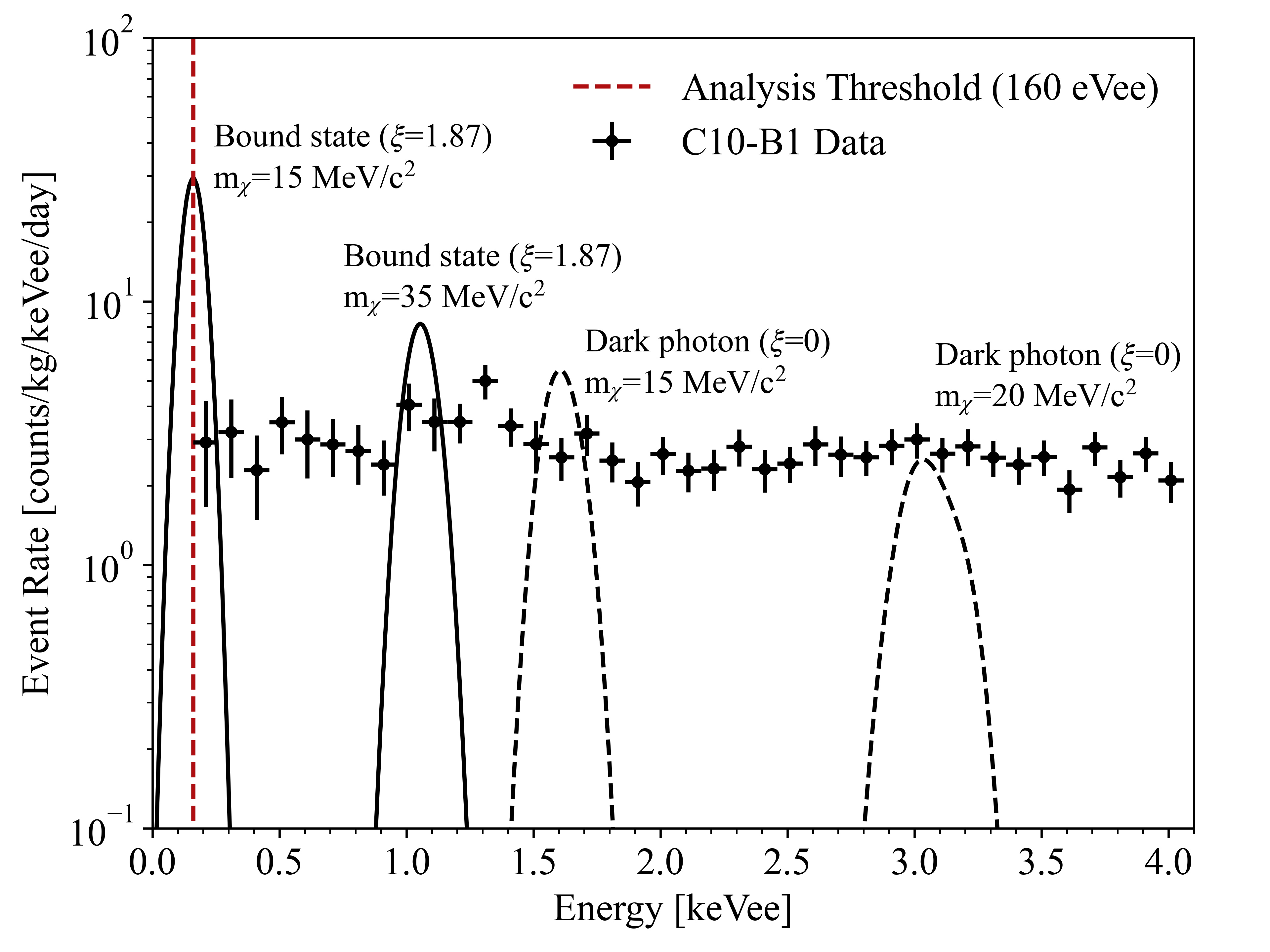}
    \caption{\label{fig:spExp3to2} Expected spectra of the 
    DM-nucleus 3$\rightarrow$2 scattering signals.
    Black dashed lines represent the spectra of the
    massless dark photon final state ($\xi=0$), 
    black solid lines denote a bound final state ($\xi=1.87$).
    The DM-nucleus coupling
    $\left \langle \sigma_{3\rightarrow 2}\cdot 
    v_\chi^2 \right \rangle n_\chi$
    is set to $10^{-45}$ cm$^2$ for both final states.
    The C10-B1 spectrum is shown as the black dots
    with the red dash line denoting the analysis threshold.}
\end{figure}

\textit{DM-nucleus 3$\rightarrow$2 scattering-}
Recent work has proposed that 
the DM may interact with the nucleus via an inelastic
3$\rightarrow$2 scattering process, 
$\chi+\chi+A\rightarrow\phi+A$,
where the DM final state ($\phi$) can be either 
a DM composite state or any dark radiation 
[\onlinecite{bib:14}]. 
The signature of this process is similar to 
that of the fermionic DM absorption. 
Neglecting the initial kinetic energy of the DM
particles, the monoenergetic nuclear recoil energy is:

\begin{equation}
    E_R\simeq\frac{(4-\xi^2)m_\chi^2}{2(M+2m_{\chi})},
    \label{eq:9}
\end{equation}

\noindent where M is the mass of the nucleus, 
$\xi$ is the mass ratio of the final and initial 
dark matter states $\phi$ and $\chi$. 
The mass ratio $\xi$ is model-dependent. 
In this study, we considered two models: 
a massless DM final state ($\phi$=dark photon, $\xi$=0) 
and a bound DM final state. 
For the bound DM final state, the mass ratio 
$\xi=2(m_\chi+\epsilon)/m_\chi$,
where the binding energy $\epsilon$ is decided 
by a new gauge coupling $g_D$:
$\epsilon=-(g_D^4\cdot m_\chi)/(64\pi^2)$ 
[\onlinecite{bib:29}].
We set the gauge coupling $|g_D|=3$ as in 
[\onlinecite{bib:14}]
and obtain a mass ratio $\xi=1.87$
for the bound DM final state.

The total event rate of the 3$\rightarrow$2 scattering 
is similar in form to that of the fermionic DM absorption:

\begin{eqnarray}
    R_{3\rightarrow 2}=
    \frac{\rho_\chi}{m_\chi}\cdot
    n_\chi \left \langle \sigma_{3\rightarrow 2}\cdot 
    v_\chi^2 \right \rangle \\ \nonumber
    \times\frac{1}{M_T}\sum_{j}^{}N_jM_jA_j^2F_j^2&&(q_{3\rightarrow 2}),
    \label{eq:10}
\end{eqnarray}

\noindent where the
$\left \langle \sigma_{3\rightarrow 2}\cdot 
v_\chi^2 \right \rangle$ 
is the average three-body inelastic cross section 
with a DM initial velocity of $v_\chi$. 
$n_\chi=\rho_\chi/m_\chi$
is the number density. 
The momentum transfer in the 3$\rightarrow$2 scattering is 
$q_{3\rightarrow 2}=\sqrt{4-\xi^2}m_\chi$.
In the calculation of DM expected spectra, 
the quenching factor and energy resolution are the same
as in fermionic DM absorption calculation. 
Fig.\ref{fig:spExp3to2} shows the expected spectra of 
the 3$\rightarrow$2 scattering signals for
the massless dark photon and
bound DM final state.
The shape of the expected spectra is also 
similar to that of the fermionic DM absorption.

\begin{figure}[!htb]
    \includegraphics
    [width=1.0\hsize]
    {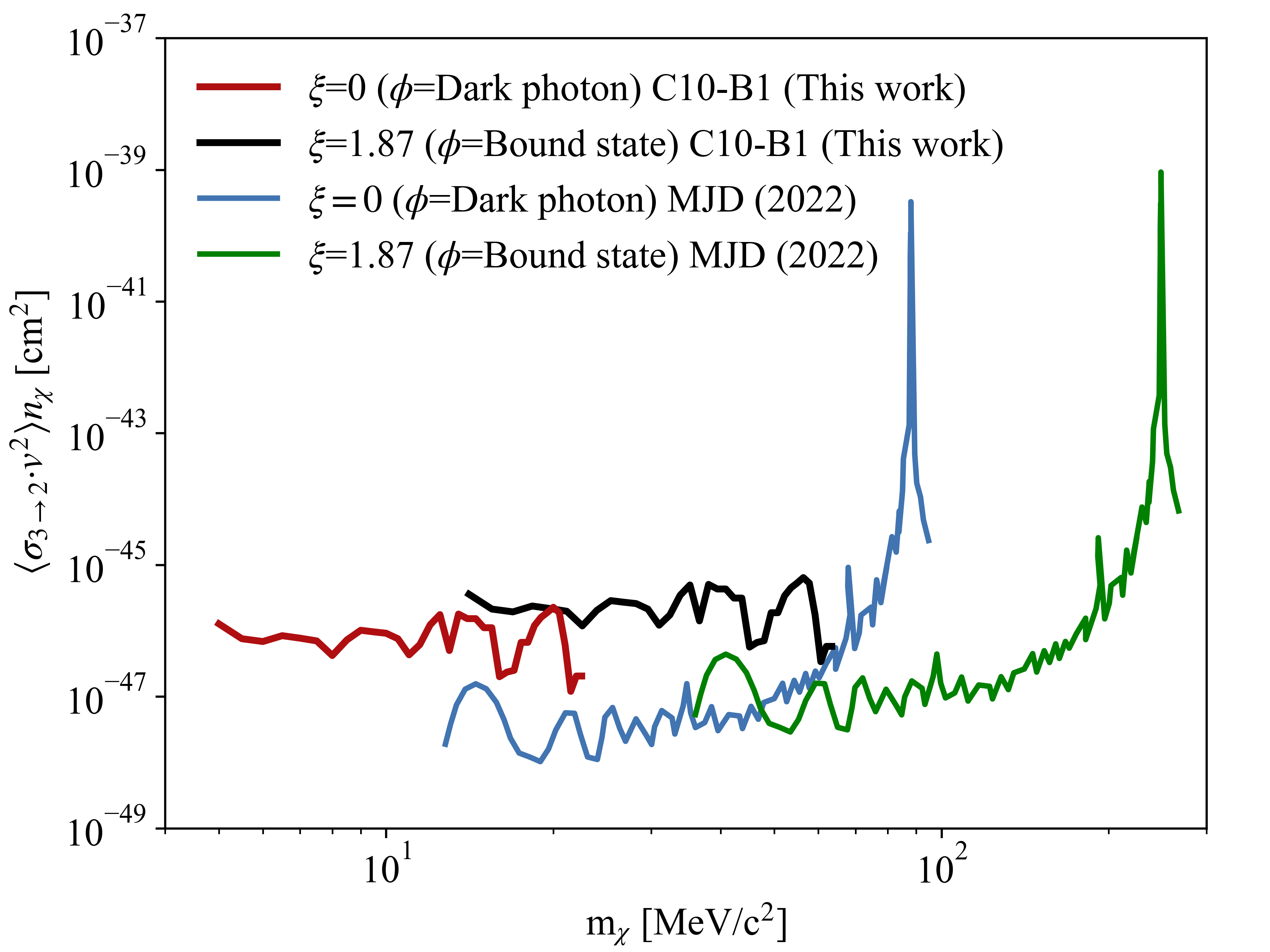}
    \caption{\label{fig:limit-3-2}Upper limit (90\% C.L.) 
    of the DM-nucleus 3$\rightarrow$2 
    scattering coupling 
    $\left \langle \sigma_{3\rightarrow 2}\cdot 
    v_\chi^2 \right \rangle n_\chi$.
    $\xi=0$ is a massless dark photon final state, 
    and $\xi=1.87$ is a bound state. 
    The results of this work (red and black lines) 
    are compared with the results from 
    the M\textsc{ajorana} D\textsc{emonstrator} 
    (MJD, blue and green lines). 
    The MJD takes a local DM density ($\rho_{\chi}$) of 0.4 GeV/cm$^3$ [\onlinecite{bib:13}],
    while this work takes a $\rho_{\chi}$ of 0.3 GeV/cm$^3$ as recommended in[\onlinecite{bib:22}].
    This work places limits on a lower DM mass region for
    both massless and bound DM final states 
    with a lower analysis threshold of 160 eVee.}
\end{figure}

We place limits on the 
$(m_\chi, \left \langle \sigma_{3\rightarrow 2}\cdot 
v_\chi^2 \right \rangle n_\chi)$
parameter as suggested in Ref [\onlinecite{bib:14}] 
for the massless dark photon and the bound DM final state. 
The 90\% upper limit for the coupling
$\left \langle \sigma_{3\rightarrow 2}\cdot 
v_\chi^2 \right \rangle n_\chi$
is shown in Fig.\ref{fig:limit-3-2}. 
Compared with the results from 
the M\textsc{ajorana} D\textsc{emonstrator} 
[\onlinecite{bib:13}], 
this work achieved a lower analysis threshold of 160 eVee 
and excluded new parameter space in lower 
DM mass regions.

\textit{Conclusion-}
In this Letter, 
we report search results for the 
sub-GeV fermionic DM absorption 
and DM-nucleus 3$\rightarrow$2 scattering
using 205.4 kg$\cdot$day exposure data from the 
C10-B1 PPCGe detector in the CDEX-10 experiment. 
The expected DM signal is analyzed in the 
160 eVee$\sim$4 keVee spectrum via a minimum $\chi^2$ method.
With an analysis threshold of 160 eVee, 
we set new experimental limits on the fermionic DM 
absorption cross section in $10\sim45$ MeV/c$^2$ DM mass. 
For the DM-nucleus 3$\rightarrow$2 scattering,
we place limit on lower DM mass of 5 MeV/c$^2$ 
for a massless dark photon DM final state,
and 14 MeV/c$^2$ for the bound DM final state.
Those limits achieve the lowest DM mass among the 
searches in direct detection experiments to date.
\\ \\ \\ \\ \\ \\

\begin{acknowledgments} 
    \setlength{\parskip}{0.0cm}
    This work was supported by the National Key Research 
    and Development Program of China 
    (Grant No. 2017YFA0402200) and 
    the National Natural Science Foundation of China 
    (Grants No. 12175112, No. 12005111, No. 11725522, No. 11675088, No.11475099).
    We would like to thank CJPL and its staff for hosting and supporting the CDEX project. 
    CJPL is jointly operated by Tsinghua University and Yalong River Hydropower Development Company.
\end{acknowledgments}

\nocite{*}


\end{document}